\documentclass[10pt,fleqn]{article}
\usepackage[utf8]{inputenc}
\usepackage[top=.8in,bottom=.8in,left=.8in,right=.8in]{geometry}
\usepackage{authblk}
\usepackage{multicol}
\usepackage{amsmath}
\usepackage{amssymb,amsfonts}
\usepackage{titlesec}
\usepackage{bbold}
\usepackage{MnSymbol}
\titleformat*{\section}{\centering\bfseries}
\newcommand{\ket}[1]{|#1\rangle}

\title{Exact Fractional Revival in Spin Chains}
\author[1]{Vincent X. Genest}
\author[1]{Luc Vinet}
\author[1,2]{Alexei Zhedanov}
\affil[1]{Centre de recherches math\'ematiques, Universit\'e de Montr\'eal, P.O. Box 6128, Centre-ville Station, Montr\'eal (QC), Canada, H3C 3J7}
\affil[2]{Donetsk Institute for Physics and Technology, Donetsk 83114, Ukraine}
\date{}
\begin{document}
\maketitle
\begin{abstract}
The occurrence of fractional revival in quantum spin chains is examined. Analytic models where this phenomenon can be exhibited in exact solutions are provided. It is explained that spin chains with fractional revival can be obtained by isospectral deformations of spin chains with perfect state transfer.
\end{abstract}
\begin{multicols}{2}
\section{Introduction}
The purpose of this Letter is to discuss fractional re\-vival in quantum spin chains and in particular to provide analytic models for which the phenomenon can be exhi\-bited in exact solutions. These systems could be used advantageously to achieve quantum information transportation and to generate entangled states. The predicted fractional revival might lend itself to experimental testing in view of the current technological advances in the simulation of spin chains with Nuclear Magnetic Resonance techniques.

Quantum wires can be thought of as devices linking parts of a quantum computer that effect the transfer of quantum states or quantum information with high fide\-lity. It has been appreciated that spin chains with engineered couplings can be quite useful in the design of quantum wires; see \cite{2007_Bose_ContempPhys_48_13,2010_Kay_IntJQtmInf_8_641, 2014_Nikilopoulos&Jex} for reviews. Central in these applications is the phenomenon of quantum revival \cite{2004_Robinett_PhysReports_392_1,2001_Berry&Marzoli&Schleich_PhysWorld_14_39}. This effect arises when a well localized wave packet spreads during its time evolution only to relocalize in its original form after some time. The transfer of states along a spin chain proceeds in this way. The chain is initialized in a state sharply peaked at one end and the dynamics is arranged so that the evolution will bring a revival of this packet at the end of the chain after finite time. To study one-qubit transfer, one looks at the one-excitation behavior. The transfer is said to be perfect when the state of the chain initially with only one spin up at the zeroth site is found with probability 1 after some time with the one spin up at the $N$\textsuperscript{th} site. The spin chains that have the property of achieving perfect state transfer (PST) and those also exhibiting almost perfect state transfer (APST) have been widely studied; in addition to \cite{2007_Bose_ContempPhys_48_13,2010_Kay_IntJQtmInf_8_641, 2014_Nikilopoulos&Jex, 2004_Albanese&Christandl&Datta&Ekert_PhysRevLett_93_230502} among the many references see for instance \cite{2005_Christandl&Datta&Dorlas&Ekert&Kay&Landahl_PhysRevA_71_032312, 2005_Yung&Bose_PhysRevA_71_032310, 2005_Shi&Li&Song&Sun_PhysRevA_71_032309, 2005_Karbach&Stolze_PhysRevA_72_030301,2012_Vinet&Zhedanov_PhysRevA_85_012323,2012_Vinet&Zhedanov_PhysRevA_86_052319,2012_Godsil&Kirkland&Severini&Smith_PhysRevLett_109_050502}. Conditions on the one-excitation spectrum are required. As a rule, the couplings between the neighboring sites as well as the local magnetic fields must be non-uniform. Algorithms based on inverse spectral theory have been elaborated to determine the specifications of those chains \cite{2010_Kay_IntJQtmInf_8_641, 2005_Yung&Bose_PhysRevA_71_032310, 2012_Vinet&Zhedanov_PhysRevA_85_012323,2012_Vinet&Zhedanov_PhysRevA_86_052319}. In particular, useful analy\-tical models that can be solved exactly have been found and seen to be related to interesting families of ortho\-gonal polynomials \cite{2004_Albanese&Christandl&Datta&Ekert_PhysRevLett_93_230502,2012_Vinet&Zhedanov_PhysRevA_85_012323,2012_Vinet&Zhedanov_JPhysConfSer_343_012125,2010_Chakrabarti&VanderJeugt_JPhysA_43_085302, 2010_Jafarov&VanderJeugt_JPhysA_43_405301}. A necessary condition for PST or APST to occur is that the Hamiltonian possesses a certain mirror symmetry, which we shall define precisely below. It is striking that an experimental quantum simu\-lation of this mirror inversion with engineered couplings could be achieved \cite{2014_Rao&Mahesh&Kumar_PhysRevA_90_012306}.

Fractional revival \cite{2004_Robinett_PhysReports_392_1, 1997_Aronstein&Stroud_PhysRevA_55_4526} occurs when a number of smaller packets, seen as little clones of the original one form at certain sites and show local periodicities. In a spin chain, one envisages that an initial state with one-excitation---a spin up---at one end would start spreading and split in transmitted and reflected parts around the middle of the chain because of interactions or impurities and would evolve after some time in a state that gives zero amplitudes to find the spin up at any of the sites except at two, or possibly a small subset of the sites. This evolution would also replicate periodically. Fractional revival of the initial packet would thus occur \cite{2007_Chen&Song&Sun_PhysRevA_75_012113}. Such dynamical phenomena for spin chains have recently been explored in \cite{2015_Banchi&Compagno&Bose_PhysRevA_91_052323} where it has been shown that the perfect revival scenario described above can be realized. 

In the following, we shall explain that spin chain Hamiltonians with fractional revival can be obtained by isospectral deformations of any system with PST. Analytic models with PST will thus lead to chains for which fractional revival can be exhibited exactly. Mirror symmetry will be traded for another symmetry in these instances. With an eye to entanglement generation, results were obtained in \cite{2010_Dai&Feng&Kwek_JPhysA_43_035302} in the special case of perfectly balanced revivals and in \cite{2010_Kay_IntJQtmInf_8_641} using chain decomposition for an odd number of sites. Our presentation offers a cohesive picture that makes clear how a properly chosen simi\-larity transformation only affects the central parameters of a chain with PST so as to produce one with fractional revival.

\section{$XX$ spin chains}
Consider $XX$ spin chains with $N+1$ sites and nearest-neighbor interactions that are governed by Hamiltonians $H$ of the form
\begin{equation}
 H=\frac{1}{2}\sum_{\ell=0}^{N-1}J_{\ell+1}(\sigma_{\ell}^{x}\sigma_{\ell+1}^{x}+\sigma_{\ell}^{y}\sigma_{\ell+1}^{y})+\frac{1}{2}\sum_{\ell=0}^{N}B_{\ell}(\sigma_{\ell}^{z}+1),
\end{equation}
where $J_{\ell}$ is the coupling constant between the sites $\ell-1$ and $\ell$ and $B_{\ell}$ is the magnetic field strength at the site $\ell$, where $\ell=0,1,\ldots, N$. The symbols $\sigma_{\ell}^{x}$, $\sigma_{\ell}^{y}$, $\sigma_{\ell}^{z}$ stand for the Pauli matrices with $\ell$ indicating that they act on the $\ell$\textsuperscript{th} $\mathbb{C}^2$ factor in $(\mathbb{C}^2)^{\otimes N+1}$. Since
\begin{align}
 \left[H,\frac{1}{2}\sum_{\ell=0}^{N}(\sigma_{\ell}^{z}+1)\right]=0,
\end{align}
the eigenstates of $H$ split in subspaces labeled by the number of spins over the chain that are up, i.e. that are eigenstates of $\sigma^{z}$ with eigenvalue $+1$, $\sigma^{z}\ket{\uparrow}=\ket{\uparrow}$. We shall call $J$ the restriction of $H$ to the subspace of states with only one spin up. The natural basis for that subspace is given by the vectors 
\begin{align}
 \ket{\ell}=(0,0,\ldots, 1,\ldots, 0),\qquad \ell=0,\ldots, N,
\end{align}
where the only ``1'' occupies the $\ell$\textsuperscript{th} entry. The action of $J$ in that basis is readily seen to be
\begin{align}
 J\ket{\ell}=J_{\ell+1}\ket{\ell+1}+B_{\ell}\ket{\ell}+J_{\ell}\ket{\ell-1},
\end{align}
with the conditions $J_0=J_{N+1}=0$ assumed; $J$ is thus given by the following Jacobi matrix
\begin{align} \label{JacobiMatrix}
 J=
 \begin{pmatrix}
B_0	&	J_1	&		&	&
\\
J_1	&	B_1	&	J_2	&	&
\\
	&	J_2	&	B_2	& \ddots &
\\
	& 		& \ddots		& 	\ddots & J_{N}
\\
 & & &J_{N}&B_{N}
 \end{pmatrix}.
\end{align}
\section{Perfect State Transfer}
Let us review the main features of PST along an $XX$ chain. Perfect transfer is achieved if at some time $T$
\begin{align}
\label{PST}
 e^{-iTJ}\ket{0}=\ket{N},
\end{align}
where a possible arbitrary phase has been set to 1. When this is so, the excitation localized at one end of the chain is transferred to the other end with probability 1 after time $T$. The main question is: for which Hamiltonians $H$ will PST occur? It is known that PST is not possible in a uniformly coupled chain, i.e. $J_{\ell}=1$ and $B_{\ell}=0$ for $\ell=0,\ldots, N$, when $N\geq 3$. What are then the coefficients $J_{\ell}$ and $B_{\ell}$ for which PST will happen? That question has been fully answered in \cite{2010_Kay_IntJQtmInf_8_641,2005_Yung&Bose_PhysRevA_71_032310, 2012_Vinet&Zhedanov_PhysRevA_85_012323}. It can be shown that a necessary condition for PST is that the matrix $J$ be mirror-symmetric with respect to the anti-diagonal: $RJR=J$ with
\begin{align}
R=
 \begin{pmatrix}
  &&&1\\
  &&1&\\
  &\udots&&
  \\
  1 &&&
 \end{pmatrix}.
\end{align}
Looking at \eqref{JacobiMatrix}, this amounts to
\begin{align}
J_{\ell}=J_{N+1-\ell}, \quad B_{\ell}=B_{N-\ell}.
\end{align}
Given that $J$ is mirror-symmetric, the additional requirements on the spectrum of $J$ are such that
\begin{align} \label{SpectrumReq}
 e^{-iTJ}=R,
\end{align}
which implies not only \eqref{PST} but a complete mirror inversion $\ket{n}\leftrightarrow \ket{N-n}$ of the register after time $T$.
\section{Fractional Revival}
We now examine the circumstances for fractional revival. We shall denote by $\widetilde{H}(\widetilde{J})$ Hamiltonians with that feature. Fractional revival happens if the initial state $\ket{0}$ evolves in a coherent sum of $\ket{0}$ and $\ket{N}$ after some time:
\begin{align}
\label{Revival}
 e^{-iT\widetilde{J}}\ket{0}=\alpha \ket{0}+\beta \ket{N},\qquad \rvert \alpha\rvert^2+\rvert \beta\rvert^2=1.
\end{align}
Simple observations allow one to see how a Hamiltonian $\widetilde{H}$ with fractional revival can be constructed from a Hamiltonian $H$ with PST. From now on we need to treat the cases where $N$ is even or odd separately.

Consider the $(N+1)\times(N+1)$ matrix $V$ defined as follows. For $N$ odd,
\begin{align}
 V=
\begin{pmatrix}
 \sin \theta &&&&&\cos \theta \\
 &\ddots &&&\udots & \\
 &&\sin \theta& \cos\theta&&\\
 &&\cos \theta& -\sin\theta &&\\
 &\udots &&&\ddots&\\
 \cos \theta &&&&&-\sin \theta
\end{pmatrix},
\end{align}
and for $N$ even
\begin{align}
 V=
\begin{pmatrix}
\sin \theta &&&&&&\cos\theta\\
& \ddots &&&&\udots&\\
&&\sin\theta&0&\cos\theta &&\\
&& 0 & 1 &0 &&\\
&&\cos \theta & 0 & -\sin\theta &&\\
&\udots &&&&\ddots&\\
\cos \theta &&&&&&-\sin\theta
\end{pmatrix},
\end{align}
with $0\leq \theta <\pi$. It is immediate to check that $V=V^{\top}$ and that $V^2=\mathbb{1}$. Furthermore, it is easy to verify that
\begin{align}
V R V=Q,
\end{align}
with $Q$ the matrix obtained from $V$, for $N$ odd and for $N$ even, by replacing $\theta$ by $2\theta$. Obviously $Q^2=\mathbb{1}$.

Let $J$ be a mirror-symmetric one-excitation Hamiltonian with the PST property. We thus have 
$e^{-iJT}=R$. Consider the ``Hamiltonian''
\begin{align}
\label{Conjugation}
\widetilde{J}=V J V.
\end{align}
It follows that
\begin{align}
e^{-iT \widetilde{J}}=Q.
\end{align}
Given the form of $Q$, we see that not only do we have \eqref{Revival} with $\alpha=\sin 2\theta$ and $\beta=\cos 2\theta$ but also
\begin{multline}
e^{-i T\widetilde{J}}\ket{\ell}=
\\
\left\{
\begin{matrix}
& \text{$N$ odd} & \text{$N$ even} \\[.2cm]
\sin 2\theta\,\ket{\ell}+\cos 2\theta\,\ket{N-\ell} & \ell\leq \frac{N-1}{2} & \ell<\frac{N}{2} \\[.1cm]
-\sin 2\theta\,\ket{\ell}+\cos 2\theta\,\ket{N-\ell} & \ell\geq \frac{N+1}{2} & \ell>\frac{N}{2}
\end{matrix}\right.,
\end{multline}
and for $N$ even $e^{-i T\widetilde{J}}\,\ket{\textstyle{\frac{N}{2}}}=\ket{\textstyle{\frac{N}{2}}}$. Such a $\widetilde{J}$, that can be obtained from any $J$ with PST, thus provides a Hamiltonian that revives a packet localized at site $\ell$ in two packets localized at sites $\ell$ and $N-\ell$. Note that the deformation that takes $J$ into $\widetilde{J}$ is iso\-spectral: the eigenvalues of $\widetilde{J}$ are the same as those of $J$. $\widetilde{J}$ is not mirror-symmetric but is rather invariant under the one-parameter involution $Q$: 
\begin{align*}
Q\widetilde{J}Q=\widetilde{J}.
\end{align*}

Interestingly, the only modifications (or perturbations) in the couplings and magnetic fields that arise in passing from $J$ to $\widetilde{J}$ occur in the middle of the chain. Working out the conjugation \eqref{Conjugation}, the only coefficients that differ are seen to be, for $N$ odd
\begin{subequations}
\label{Perturbations}
\begin{align}
\begin{aligned}
\widetilde{J}_{\frac{N+1}{2}}&=J_{\frac{N+1}{2}}\,\cos 2\theta,
\\
\widetilde{B}_{\frac{N\mp 1}{2}}&= B_{\frac{N-1}{2}}\pm J_{\frac{N+1}{2}}\sin 2\theta,
\end{aligned}
\end{align}
and for $N$ even
\begin{align}
\begin{aligned}
\widetilde{J}_{\frac{N}{2}}&= J_{\frac{N}{2}}(\cos \theta+\sin \theta),
\\
\widetilde{J}_{\frac{N}{2}+1}&=J_{\frac{N}{2}}(\cos \theta-\sin \theta).
\end{aligned}
\end{align}
\end{subequations}
Recall that $J$ is mirror-symmetric. When $N$ is even, only the couplings between the 3 middle neighbors are altered. When $N$ is odd, only the coupling  between the 2 middle neighbors is affected as well as the magnetic field strengths at those two middle sites. Note that if all the $B_{\ell}$ were initially zero in $J$, there would only be two Zeeman terms of equal magnitude and opposite sign at $\ell=\frac{N-1}{2}$ and $\ell=\frac{N+1}{2}$. Observe also that the para\-meter $\theta$ entering in $\widetilde{J}$ and hence in $\widetilde{H}$ directly determines the probability of finding at time $T$ the spin up at the input or output site. Clearly, analytic spin chain models with fractional revival can thus be obtained directly from the analytic models with PST that have already been identified. The paradigm example is associated to the Krawtchouk orthogonal polynomials and has
\begin{align} \label{KrawtchoukCoefficients}
\begin{aligned}
B_{\ell}=0,\;
J_{\ell}=\frac{1}{2}\sqrt{\ell(N+1-\ell)},\; \ell=0,1,\ldots, N.
\end{aligned}
\end{align}
It is readily checked that \eqref{SpectrumReq} is obeyed with these couplings for $T=\pi$.
Modifying the coefficients \eqref{KrawtchoukCoefficients} according to \eqref{Perturbations} will provide a most simple system where fractional revival can be identified exactly.

In \cite{2015_Banchi&Compagno&Bose_PhysRevA_91_052323}, the authors have followed an approach that differs from the one presented here. They have considered chains that preserve mirror symmetry. They have determined numerically in one case the specifications of such a chain with fractional revival, using the parameters of the Krawtchouk chain as initial conditions. In this example, contrary to the isospectral deformations, all the couplings of the chain are perturbed. As a matter of fact, an exact solution already existed \cite{2012_Vinet&Zhedanov_JPhysA_45_265304} for this model that has the property of exhibiting both fractional revival and PST. More details will be given in a forthcoming publication.
\section{Conclusion}
Let us conclude with remarks on applications. Spin chains with fractional revival can be used to transfer quan\-tum information. Assume that information entered at the beginning of the chain is to be transported to the end of the chain where it will be processed in a quantum computation. By adjusting the parameter $\theta$ of the chain, one can arrange that this information is at site $N$ with probability bigger than $1/2$ at precise periodical times. Upon repeating the calculation using the information at the end of the chain at those times, the likelihood that the most frequent answer is right will be high. Let us also stress that these spin chains with revival at two sites could prove to be an interesting tool to generate entangled states. Indeed it is immediate to see that when the revival is balanced ($\theta=\pi/8$) the sites $0,1,N-1$ and $N$ for instance would support the entangled state $\frac{1}{\sqrt{2}}\left(\ket{\uparrow \downarrow}+\ket{\downarrow \uparrow}\right)$.
\section*{Acknowledgments} 
The authors would like to thank A. Kay and S. Severini for bringing \cite{2010_Dai&Feng&Kwek_JPhysA_43_035302} to their attention. They are also grateful to L. Banchi, S. Bose, M. Christandl, G. Coutinho and S. Severini for stimulating exchanges and for sharing their advances prior to publication. V.X.G. holds a fellowship from the Natural Science and Engineering Research Council of Canada (NSERC). The research of L.V. is supported in part by NSERC. A.Z. wishes to thank the Centre de recherches math\'ematiques (CRM) for its hospitality.
\small

\end{multicols}
\end{document}